\input{aipcheck}

\documentclass[
    ,final            
  ]
  {aipproc}

\layoutstyle{6x9}

\begin{document}

\title{Bound States of (Anti-)Scalar-Quarks \\ in $\bf{SU(3)_c}$ Lattice QCD}

\author{H. Iida}{
  address={Yukawa Institute for Theoretical Physics, Kyoto University, 
Sakyo, Kyoto 606-8502, Japan}
}

\author{H. Suganuma}{
  address={Department of Physics, Graduate School of Science, Kyoto University, 
Kyoto 606-8502, Japan
}
}

\author{T.T. Takahashi}{
  address={Yukawa Institute for Theoretical Physics, Kyoto University, 
Sakyo, Kyoto 606-8502, Japan}
}

\begin{abstract}
Light scalar-quarks $\phi$ (colored scalar particles or idealized diquarks) and 
their color-singlet hadronic states are studied 
with quenched SU(3)$_c$ lattice QCD in terms of mass generation. 
We investigate ``scalar-quark mesons'' $\phi^\dagger \phi$ and ``scalar-quark baryons'' $\phi\phi\phi$ 
as the bound states of scalar-quarks $\phi$.
We also investigate the bound states of scalar-quarks $\phi$ and quarks $\psi$, i.e., 
$\phi^\dagger \psi$, $\psi\psi\phi$ and $\phi\phi\psi$, which we name ``chimera hadrons''. 
All the new-type hadrons including $\phi$ are found to have a large mass 
due to large quantum corrections by gluons, 
even for zero bare scalar-quark mass $m_\phi=0$ at $a^{-1}\sim 1{\rm GeV}$. 
We conjecture 
that all colored particles generally acquire 
a large effective mass due to dressed gluon effects.
\end{abstract}

\maketitle



\section{Mass Generation of Colored Particles in QCD}
The origin of mass is a fundamental and fascinating subject in physics for a long time. 
A standard interpretation of mass origin is the interaction with the Higgs field. 
However, the mass of Higgs origin is only about 1\% of total mass in the world, 
because the Higgs interaction only provides  
the current quark mass (less than 10MeV for light quarks) 
and the lepton mass (0.51MeV for electrons).
On the other hand, apart from unknown dark matter, 
about 99\% of mass of matter in the world originates from the strong interaction, 
which actually provides the large constituent quark mass $M_\psi =(300-400){\rm MeV}$. 

Dynamical fermion-mass generation in the strong interaction can be 
interpreted as spontaneous breaking of chiral symmetry, 
which was first pointed out by Y.~Nambu~et~al. \cite{NJ61}. 
According to the chiral symmetry breaking, light quarks are considered to have 
a large constituent quark mass of about $400{\rm MeV}$. 

Then, a question arises. Is there any mechanism of dynamical mass generation without 
chiral symmetry breaking? 
To answer this question, we note the following examples.
One example is gluons, colored vector particles in QCD.
While the gluon is massless in perturbation QCD, 
non-perturbative effects due to the self-interaction of gluons seem to generate 
a large effective gluon mass as $(0.5-1.0){\rm GeV}$, which is measured in lattice QCD \cite{MO87,AS99}. 
Actually, glueballs, which are composed by gluons, 
have a large mass, e.g., about $1.5{\rm GeV}$ \cite{MP99ISM02} 
even for the lightest glueball ($J^{PC}=0^{++}$). 
Another example 
is charm quarks. The current mass of charm quarks is about $1.2{\rm GeV}$ 
at the renormalization point $\mu=1{\rm GeV}$ \cite{PDG}. 
In the quark model, however, the constituent charm-quark mass is about $1.6{\rm GeV}$, 
which reproduces masses of charmonia \cite{RGG75}. 
The about 400MeV difference between the current and the constituent charm-quark masses 
can be explained as dynamical mass generation without chiral symmetry breaking,
since there is no chiral symmetry for such a heavy-quark system. 

These examples suggest that there is other type of mass generation 
without chiral symmetry breaking and the Higgs mechanism. 
Then, we conjecture that, even without chiral symmetry breaking, large dynamical mass generation 
generally occurs in the strong-interaction world, i.e., 
{\it all colored particles 
have a large effective mass generated by dressed gluon effects.}

\section{Scalar-Quarks and Chimera Hadrons in Lattice QCD}

In this paper, we study light (${\bf 3}_c$-colored) ``scalar-quarks'' $\phi$ 
and their color-singlet hadronic states 
using quenched SU(3)$_c$ lattice QCD.
This is a test-place for dynamical mass generation without chiral symmetry breaking. 
Also, this light scalar-quark can be regarded as ``diquark'' idealized to be  
a local field at the scale of $a^{-1}\sim 1{\rm GeV}$, 
since the diquark picture is often used as an important degrees of freedom in hadron physics. 

Here, we investigate ``scalar-quark mesons'' $\phi^\dagger \phi$ 
and ``scalar-quark baryons'' $\phi\phi\phi$
as the bound states of scalar quarks $\phi$. 
We also investigate bound states of scalar-quarks $\phi$ and quarks $\psi$, i.e., 
$\phi^\dagger \psi$, $\psi\psi\phi$ and $\phi\phi\psi$, which we name ``chimera hadrons''. 

To include scalar-quarks $\phi$ together with quarks $\psi$ and gluons in QCD,  
we adopt the generalized QCD Lagrangian density, 
\begin{eqnarray}
{L}= -\frac{1}{4}G^a_{\mu\nu}G^{a\mu\nu} + {L}_{\rm F} + {L}_{\rm SQ}, 
\ \ \ {L}_{\rm SQ}={\rm tr} \ (D_\mu \phi)^\dagger(D^\mu\phi)-m_{\phi}^2 \ {\rm tr} \ \phi^\dagger\phi,
\end{eqnarray}
where ${L}_{\rm F}$ denotes the quark part and
$m_\phi$ the bare mass of scalar-quarks $\phi$. 
In the lattice formalism, we adopt the Euclidean lattice action 
for the scalar-quark sector as
\begin{eqnarray}
S_{\rm SQ}\equiv \sum_{x,y}{\phi}^\dagger(x)\left\{
-\sum_\mu(\delta_{{x+\hat\mu},y}U_\mu (x)+\delta_{{x-\hat\mu},y}
{U_\mu}^\dagger(y)-2\delta_{xy}{\bf 1})+m_{\phi}^2\delta_{xy}{\bf 1}
\right\} {\phi}(y),
\end{eqnarray}
where $U_\mu(x)$ is the link-variable. 
For the gluon sector, we adopt the standard Wilson action with 
$\beta\equiv\frac{2N_c}{g^2}=5.7$, i.e., 
$a^{-1} \simeq 1.1{\rm GeV}$ on the lattice spacing \cite{T0102}. 
For the bare scalar-quark mass, we take $m_{\phi}$=0.0, 0.11, 0.19 and 0.26GeV. 
For the quark sector, we adopt the Wilson quark action with 
the bare quark mass of $m_{\psi}$=0.09, 0.14 and 0.19GeV. 
The lattice QCD calculations have been performed on the supercomputer NEC-SX5 at Osaka University 
with the setup summarized in Table~1. 

\begin{table}[b]
\begin{tabular}{cccccc}
\hline
$\beta$ &Lattice size& $a^{-1}$&
bare scalar-quark mass $m_{\phi}$ & bare quark mass $m_{\psi}$\\ 
\hline
$5.7$ &$16^3\times 32$&  1.1 GeV
& 0.00, 0.11, 0.19, 0.26GeV
&0.09, 0.14, 0.19GeV\\
\hline
\end{tabular}
\caption{The lattice QCD setup for scalar-quark hadrons and chimera hadrons.}
\end{table}

\begin{table}[ht]
\begin{tabular}{lll}
\hline
Scalar-quark meson ($\phi^\dagger\phi$) & (scalar) & $M_s(x)\equiv \phi^\dagger_a (x)\phi_a(x)$\\
Scalar-quark baryon ($\phi\phi\phi$) & (scalar) & $B_s(x)\equiv \epsilon_{abc}\phi_a (x)\phi_b (x)\phi_c (x)$\\
Chimera meson ($\phi^\dagger\psi$) & (spinor) & $C_M^\alpha(x)\equiv \phi^\dagger_a (x) \psi_a^\alpha (x)$\\
Chimera baryon ($\psi\psi\phi$) & (scalar) & $C_{B}(x)\equiv \epsilon_{abc} ({\psi_a}^T(x)C\gamma_5\psi_b(x)) \phi_c (x)$\\
Chimera baryon ($\phi\phi\psi$) & (spinor) & $C_{B}^{\alpha}(x)\equiv \epsilon_{abc}\phi_a (x)\phi_b (x)\psi_c^\alpha (x)$\\
\hline
\end{tabular}
\caption{Operators of scalar-quark hadrons and chimera hadrons. 
Some of the new-type hadrons have different statistics from ordinary hadrons. 
For instance, the scalar-quark baryon $\phi\phi\phi$ 
is a boson and the chimera meson $\phi^\dagger \psi$ is a fermion. 
}
\end{table}

The gauge-invariant hadronic operators $O(\vec x,t)$ 
of scalar-quark hadrons and chimera hadrons are summarized in Table~2.
We calculate in lattice QCD the temporal correlator 
$G(t)\equiv \frac{1}{V}\sum_{\vec x} \langle O(\vec x,t) O^{\dagger}(\vec 0,0)\rangle$, 
where the total momentum is projected to be zero.
The mass $M$ of scalar-quark/chimera hadrons are obtained as 
$M \simeq -\frac{1}{T}{\rm ln} G(T)$ for large $T$.

Now, we show lattice results for the mass of scalar-quark hadrons, 
which are composed only by (anti-)scalar-quarks. 
Figure 1 shows scalar-quark meson masses $M_{\phi^\dagger \phi}$ 
obtained from connected diagrams, 
and scalar-quark baryon masses $M_{\phi\phi\phi}$, 
plotted against the bare scalar-quark mass $m_{\phi}$ at $a^{-1} \simeq 1.1{\rm GeV}$.
Even for zero bare scalar-quark mass, $m_{\phi}=0$,
scalar-quark hadrons have a large mass as 
$M_{\phi^\dagger\phi} \simeq 3{\rm GeV}$ and $M_{\phi\phi\phi} \simeq 4.7{\rm GeV}$.

\begin{figure}[ht]
\begin{minipage}{.45\linewidth}
\rotatebox{-90}{
\includegraphics[width=4.5cm]{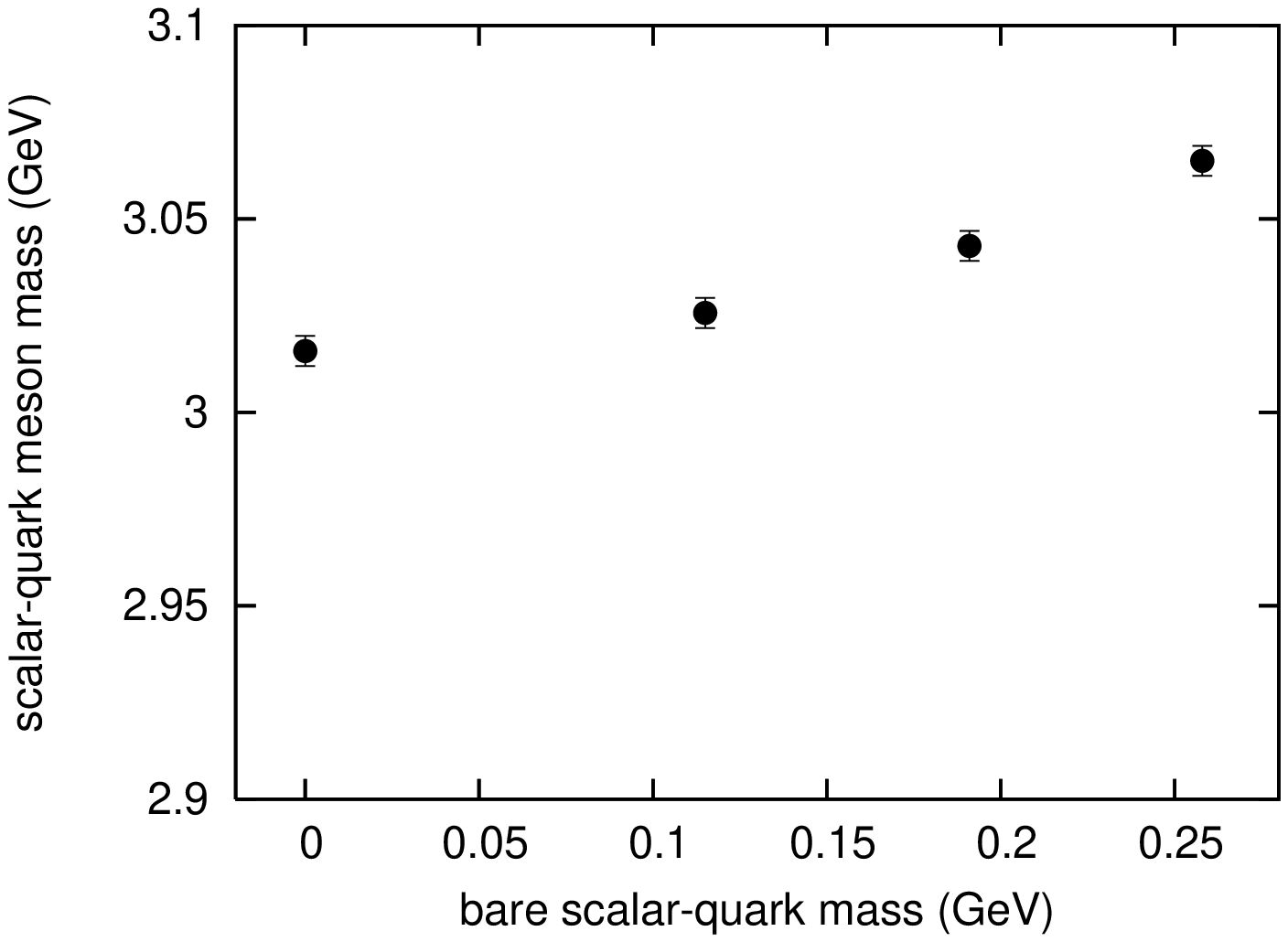}}
\label{fig1}
\end{minipage}
\hspace{1cm}

\begin{minipage}{.45\linewidth}
\rotatebox{-90}{
\includegraphics[width=4.5cm]{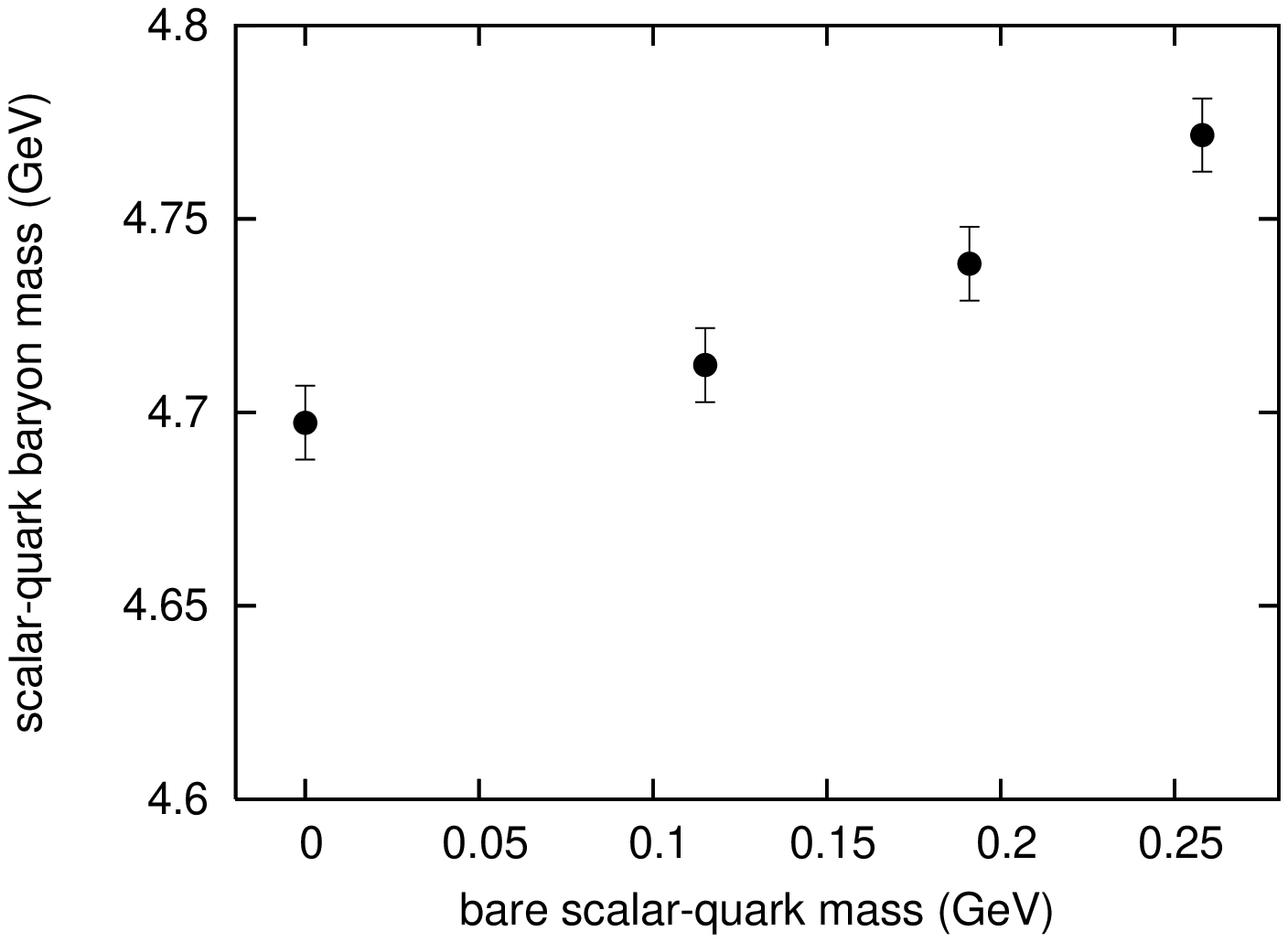}}
\caption{``Scalar-quark hadron'' masses in lattice QCD:
(a) ``scalar-quark meson'' masses $M_{\phi^\dagger \phi}$ and 
(b) ``scalar-quark baryon'' masses $M_{\phi\phi\phi}$, 
plotted against the bare scalar-quark mass $m_{\phi}$ 
at $a^{-1} \simeq 1.1{\rm GeV}$.
}
\label{fig2}
\end{minipage}
\end{figure}

Next, we show lattice QCD results for the mass of chimera hadrons, 
which are composed by scalar-quarks $\phi$ and quarks $\psi$. 
Table 3 shows chimera hadron masses 
in terms of the bare scalar-quark mass $m_\phi$ and the bare quark mass $m_\psi$ 
at $a^{-1} \simeq 1.1{\rm GeV}$.
Even near $m_\phi=m_\psi=0$, 
we find large mass generation as $M_{\phi^\dagger \psi} \simeq 1.9{\rm GeV}$ 
for chimera mesons $\phi^\dagger\psi$, 
and $M_{\psi\psi\phi} \simeq 2.2{\rm GeV}$, $M_{\phi\phi\psi} \simeq 3.6{\rm GeV}$ 
for chimera baryons ($\psi\psi\phi$, $\phi\phi\psi$).

For scalar-quark/chimera hadrons at $m_\phi=m_\psi=0$, we find 
an approximate relation as $M(n_1 \phi n_2 \psi) \simeq n_1 M_{\phi} + n_2 M_{\psi}$  
with the constituent quark mass $M_{\psi}\simeq 0.4{\rm GeV}$, and thus obtain  
a large constituent scalar-quark mass $M_{\phi}\simeq 1.5-1.6{\rm GeV}$ 
dynamically generated.

\begin{table}[ht]
\caption{The mass of chimera hadrons 
(bound states of scalar-quarks $\phi$ and quarks $\psi$) 
in term of the bare scalar-quark mass $m_{\phi}$ and the bare quark mass $m_\psi$ 
at $a^{-1} \simeq 1.1{\rm GeV}$.
$M_{\phi^\dagger\psi}$, $M_{\psi\psi\phi}$, $M_{\phi\phi\psi}$
denote the mass of chimera mesons $\phi^\dagger\psi$ and 
chimera baryons ($\psi\psi\phi$, $\phi\phi\psi$).
The mass unit is GeV. \newline 
*The values at $m_\psi=0.0$ are obtained by the linear extrapolation of $m_\psi$. 
}
\label{table1}
\begin{tabular}{cccccc}
\hline
$m_{\phi}$&$m_\psi$ & $M_{\phi^\dagger\psi}$ & $M_{\psi\psi\phi}$ & $M_{\phi\phi\psi}$\\
\hline
  \ \ \ \ \ \ 0.00 \ \ \ \ \ \ 
& \ \ \ \ \ \ 0.0* \ \ \ \ \ \ 
& \ \ \ \ \ \  1.85 \ \ \ \ \ \ \ \ \ \ \ \ \ \ \ \ \ \ \ \  
& \ \ \ \ \ \  2.21 \ \ \ \ \ \ \ \ \ \ \ \ \ \ \ \ \ \ \ \ \ 
& \ \ \ \ \ \  3.55 \ \ \ \ \ \ \ \ \ \ \ \ \ \ \ \ \ \ \ \ \ \\
0.00 & 0.09& 1.914$\pm 0.008$& 2.366$\pm 0.022$& 3.607$\pm 0.017$\\
0.00 & 0.14& 1.950$\pm 0.007$& 2.464$\pm 0.018$& 3.637$\pm 0.014$\\
0.00 & 0.19& 1.986$\pm 0.006$& 2.558$\pm 0.016$& 3.672$\pm 0.013$\\

0.11 & 0.09& 1.920$\pm 0.008$& 2.371$\pm 0.022$& 3.617$\pm 0.017$\\
0.11 & 0.14& 1.955$\pm 0.007$& 2.470$\pm 0.018$& 3.647$\pm 0.014$\\
0.11 & 0.19& 1.991$\pm 0.006$& 2.564$\pm 0.016$& 3.682$\pm 0.013$\\

0.19 & 0.09& 1.929$\pm 0.008$& 2.381$\pm 0.022$& 3.635$\pm 0.017$\\
0.19 & 0.14& 1.964$\pm 0.007$& 2.479$\pm 0.018$& 3.665$\pm 0.014$\\
0.19 & 0.19& 2.000$\pm 0.006$& 2.573$\pm 0.016$& 3.700$\pm 0.013$\\

0.26 & 0.09& 1.941$\pm 0.008$& 2.393$\pm 0.023$& 3.658$\pm 0.017$\\
0.26 & 0.14& 1.976$\pm 0.007$& 2.491$\pm 0.018$& 3.688$\pm 0.014$\\
0.26 & 0.19& 2.012$\pm 0.006$& 2.585$\pm 0.016$& 3.723$\pm 0.013$\\
\hline

\end{tabular}
\end{table}

The large mass generation of scalar-quarks $\phi$ reflects 
large quantum corrections for scalar particles, 
similar to the necessity of fine tuning for Higgs scalar fields.
This lattice result also indicates that 
the simple modeling which treats the diquark as a local scalar field 
at the scale of $a^{-1} \sim 1{\rm GeV}$ in QCD is 
rather dangerous and needs subtle fine tuning.

To summarize, we have studied light scalar-quarks $\phi$ 
(colored scalar particles or idealized diquarks) 
and their color-singlet hadronic states in quenched SU(3)$_c$ lattice QCD 
in terms of dynamical mass generation. 
We have investigated the mass of ``scalar-quark mesons'' $\phi^\dagger \phi$, 
``scalar-quark baryons'' $\phi\phi\phi$ and ``chimera hadrons''  
$\phi^\dagger \psi$, $\psi\psi\phi$, $\phi\phi\psi$, 
which are composed by quarks $\psi$ and scalar-quarks $\phi$. 
For these new-type hadrons including $\phi$, we have observed 
large dynamical mass generation of scalar-quarks as $M_{\phi} \simeq 1.5-1.6{\rm GeV}$ 
due to large quantum corrections by gluons, 
even for zero bare scalar-quark mass $m_\phi=0$ at $a^{-1}\simeq 1.1{\rm GeV}$. 

In this way, even without chiral symmetry breaking, 
large dynamical mass generation occurs for the scalar-quark systems 
as scalar-quark/chimera hadrons.
Together with the large glueball mass ($>$ 1.5GeV) and 
large difference ($\sim$ 400MeV) between current and constituent charm-quark masses, 
this type of mass generation would be generally occurred in the strong interaction, 
and therefore we conjecture that all colored particles generally acquire 
a large effective mass due to dressed gluon effects as shown in Fig.2.

\begin{figure}[ht]
\includegraphics[width=13cm]{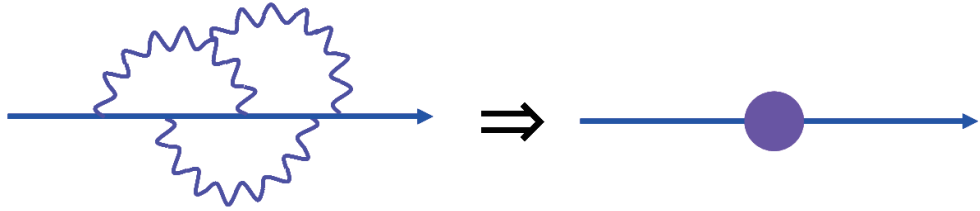}
\caption{Schematic figure for dynamical mass generation of colored particles. 
Even without chiral symmetry breaking, colored particles generally acquire 
a large effective mass due to dressed gluons.}
\label{fig2}
\end{figure}






\end{document}